\documentclass[12pt]{iopart}
\usepackage{iopams}
\usepackage{harvard}
\usepackage{color}
\usepackage{url}

\expandafter\let\csname equation*\endcsname\relax
\expandafter\let\csname endequation*\endcsname\relax
\usepackage{amsmath}
\usepackage{graphicx}
\usepackage[usenames,dvipsnames]{xcolor}
\usepackage{multirow}

\begin{document}

\title[Radiative lifetimes and cooling functions for astrophysically
important molecules]{Radiative lifetimes and cooling functions for astrophysically important molecules}

\author{Jonathan Tennyson$^{1,2}$, Kelsey Hulme$^{1}$, Omree K. Naim$^{1}$ and Sergei N. Yurchenko$^{1}$}

\address{$^1$Department of Physics \& Astronomy, University College London,
Gower St.,
London, WC1E 6BT, UK}

\ead{$^2$j.tennyson@ucl.ac.uk}

\begin{abstract}

  Extensive line lists generated as part of the ExoMol project are
  used to compute lifetimes for individual rotational, rovibrational
  and rovibronic excited states, and temperature-dependent cooling
  functions by summing over all dipole-allowed transitions for the
  states concerned. Results are presented for SiO, CaH, AlO, ScH,
  H$_2$O and methane. The results for CH$_4$ are particularly unusual
  with 4 excited states with no dipole-allowed decay route and
  several others where these decays lead to exceptionally long
  lifetimes.  These lifetime data should be useful in models of masers
  and estimates of critical densities, and can provide a link with
  laboratory measurements.  Cooling functions are important in stellar
  and planet formation.

\end{abstract}

\maketitle

\section{Introduction}

There are vast areas of the Universe where molecules exist at
very low temperatures. However molecules actually occur in a wide variety of
environments many of them significantly hotter including planetary nebulae,
atmospheres of (exo-)planets, brown dwarfs and cool stars. This makes
the radiative and cooling properties of the molecules important for models
of these species. Indeed, although interstellar molecular clouds are
usually characterised as cold, they are not really fully thermalised.
Whether a species, or state of that species, is thermalised depends
on the critical density which is given by the ratio of the radiative
lifetime of the state to the rate for collisional excitation to the state.
In such regions radiative lifetimes are also important for models
of the many species which are observed to mase.
The long lifetimes associated with certain excited states can lead
to population trapping and non-thermal distributions. Such behaviour
has been observed for the H$_3^+$ molecule both
in space \cite{02GoMcGe.H3+,05OkGeGo.H3+} and the laboratory
\cite{jt306,jt340}, in both cases leading to state distributions
which had not been anticipated.

States with short radiative lifetimes are important for direct laser cooling of polar
diatomic molecules, a topic which  has recently attracted significant interest \cite{10ShBaDe.SrF}.

Cooling functions are important in a variety of environments. For
example cooling by H$_3^+$ is known to be important in the atmospheres
of solar system gas giants \cite{jt498} and, it has been suggested,
vital to stability of so-called ``hot Jupiter'' exoplanets
\cite{kam07}.  Similarly, cooling functions are important in the
primordeal Universe for the formation of the first stars \cite{abel}
and galaxies \cite{b20,bromm}. At temperatures below about 8000 K, this
cooling is almost entirely provided by molecules \cite{jt489}. Although
cooling functions are available for some key diatomics \cite{jt506} and
there have been several attempts to provide cooling functions
for H$_3^+$ \cite{jt181,jt489,jt551}, there are no systematic
compilations of these. Indeed cooling functions for many key species,
such as water, do not appear to be available.

The ExoMol project \cite{jt528} has been undertaking the systematic
calculation of very extensive spectroscopic line lists for molecules
deemed to be important for the study of  exoplanet and other hot atmospheres.
These line lists provide comprehensive datasets of a molecule's
radiative properties. Here we use these datasets to systematically
study radiative lifetimes of comprehensive sets of states and to compute
temperature-dependent cooling functions for molecules considered as part of the
ExoMol project.

\section{Theory}

The results of ExoMol calculations have, up until now, been stored in
two files \cite{jt548}: a states file comprising one, numbered line
per state which gives the energy level of the state plus associated
quantum numbers, and a transitions file with gives a list of Einstein
A coefficients, $A_{if}$, plus the numbers of the initial, $i$, and
final, $f$, states which, in turn, refer back to the states file.
The lifetime of state $i$, $\tau_i$, can be computed by summing
over A coefficients:
\begin{equation}
\tau_i =  \frac{1}{\sum_f A_{if}}~.
\end{equation}
This is in principle straightforward, although the sheer size of some of the
transitions files, which can contain many billions of entries
\cite{jt564,jt592,jt614}, does mean that the data handling requires some care.

At temperature $T$, the cooling function, $W(T)$, is the total energy emitted by a molecule and is
given by
\begin{equation}
W(T) = \frac{1}{4 \pi Q(T)} \sum_{i,f} A_{if} h c \nu_{if} (2J_i+1) g_i \exp\left(\frac{-c_2 \tilde{E}_i}{T}\right)
\end{equation}
where $\nu_{if}$ is the wavenumber of the transition $i \to f$,
$J_i$ is the rotational quantum number of the initial state and $g_i$ is its nuclear
spin degeneracy factor. The final, exponential term is the Boltzmann factor
for which we use the second radiation constant, $c_2 = 1.438 777 36$ cm K, since ExoMol
actual stores energies $\tilde{E}$ as wavenumbers in cm$^{-1}$. Finally, $Q(T)$
is the partition function given by
\begin{equation}
Q(T) = \sum_i (2J_i+1)  g_i \exp\left(\frac{-c_2 \tilde{E}_i}{T}\right).
\end{equation}
Partition functions are computed routinely for each molecule
studied by the ExoMol project.

The general procedure used by ExoMol to produce linelists is to use
potential energy curves and surfaces obtained by refining high
quality {\it ab initio} curves and surfaces using spectroscopic data. Wavefunctions
are obtained using variational nuclear motion codes \cite{jt338,lr07,07YuThJe.method,jt588,jt609}.
Conversely dipole moment curves and surfaces are obtained  {\it ab initio}:
experience has shown that, at least for molecules containing light
atoms, these are capable of giving results as accurate as the best
measurements \cite{jt509,jt613}.

\section{Results: lifetimes}

Lifetimes were obtained for molecules treated by the ExoMol project
\cite{jt578}. Examples are analysed below with complete sets of data
being placed on the ExoMol website (www.exomol.com). While the ExoMol
line lists are very extensive, with a few exceptions
\cite{jt347,jtpoz}, they are not complete.  Instead the project aims
to provide comprehensive line lists which are effectively complete for a given
upper temperature and range of wavelengths. These choices vary with
the expected stability of a given species but typical objectives are the
provision of a
line list which represents all absorbtion longwards of 1 $\mu$m (at
wavenumbers below 10~000 cm$^{-1}$) for temperatures up to 3000~K.
In general these parameter ranges are increased for diatomics species which may be
stable in environments as hot as
our Sun's photosphere (almost 6000 K) and for which absorptions at visible
wavelengths are of importance. Conversely, for large molecules such
as nitric acid (HNO$_3$) the range of temperatures considered
is significantly reduced, to only 500 K in this case \cite{jt614}. These
choices are partially influenced by practical considerations as calculations
on polyatomic systems become very  computationally expensive \cite{jt564}.
These considerations influence the states for which lifetime data is
presented below.

By definition the ground state of all systems considered has an infinite
lifetime to radiative decay. Some excited states also do not have
allowed radiative decay routes, for example where these are forbidden
by nuclear spin considerations. In principle, radiative nuclear spin
transitions which interchange ortho and para nuclear spin isomers can
occur but they are extremely weak \cite{jt329} and such
transitions are not considered
as part  of ExoMol. States whose lifetimes are computed to
be infinite are not considered below. In the following subsections
we present lifetime results for a series of molecules starting with
diatomics, ordered by the complexity of their electronic structure,
and followed by the polyatomic molecules water and methane.

\subsection{SiO}

The ExoMol project has provided line lists for a number of closed
shell diatomic species including NaCl and KCl \cite{jt563}, PN \cite{jt590},
NaH \cite{jt605}, CS \cite{jt615} and CaO \cite{jt618}. For NaH and
CaO these studies also considered electronically excited state. Here
we consider SiO as an example. The silicon monoxide linelists of \citeasnoun{jt563} are
particularly comprehensive, considering all vibrational  and rotational level up
to the electronic ground state, corresponding to 
$v \leq 98$ and $J \leq 423$. They are  expected to be complete for
temperatures up to 9000 K and considered wavelenghts much shorter than
1 $\mu$m. We note that these very weak, short wavelength transitions
are likely to be subject to numerical problems \cite{15LiGoRo.CO,highv}
which, however, are not expected to influence the results presented below.
In addition, SiO is a well-studied maser whose emissions have been
observed in a variety of sources and at several wavelengths
\cite{99Gray,04DeFuGl.SiO,08McHaxx.SiO,09CoChKi.SiO,10LiAnSh.SiO,10CoRaP3.SiO}.

\begin{figure}[htb!]
 \centering
\includegraphics[width=8.3cm]{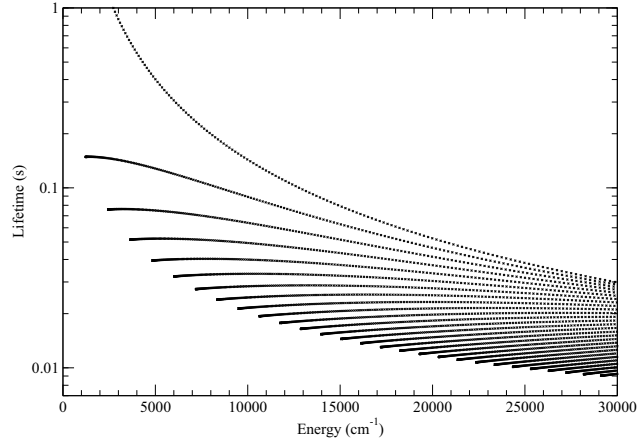}
\vspace{0.5cm}
 \caption{Lifetimes of the vibration-rotation states of $^{28}$Si$^{16}$O.
From the top
the curves are for vibrational state $v=0$, $v=1$ up to $v=27$. Curves
comprise states of increasing rotational quantum number, $J$, with energy.
State with $J \leq 61$ for $v=0$ are too long-lived to fit in the figure
and are discussed in the text.}
  \label{f:SiO}
\end{figure}

Figure~\ref{f:SiO} gives lifetimes for vibration-rotation states of
 $^{28}$Si$^{16}$O. In general these lifetimes vary smoothly with vibrational
state, $v$, and rotational state, $J$, showing reduced lifetimes upon
excitation of either mode. States in the $v=0$ vibrational ground state
can only decay by a single, $\Delta J =1$, transition. The rigid rotor
approximation suggests that such decays should lead to lifetimes which
depend on $J^{-3}$
and this behaviour is followed by our results. This means that the first
excited state with $J=1$ has a long lifetime, which we compute to be
$3.32 \times 10^5$ s.

The lifetimes of the vibrational-rotation states of other closed shell diatomic
molecules we studied behave in a similar fashion.

\subsection{CaH}

ExoMol has considered the vibration-rotation spectrum
for a number of molecules
whose electronic state also contains angular momentum \cite{jt529}.
In this case it
is necessary to consider a variety of couplings which are not present
in a non-relativistic, Born-Oppenheimer treatment of the problem such
as spin-orbit, spin-spin and spin-rotation coupling. Here we consider
calcium mononhydride, CaH, as an example.

The electronic ground state of calcuim monohydride has $^2\Sigma^+$ symmetry. A line list
for it was computed by \citeasnoun{jt529} with $v \leq 19$ and $J \leq 74$. 
We note that an alternative
line list, which also considered transitions to the first 5 electronic
states, has been provided by \cite{03WeStKi.CaH}; this line list,
which is purely {\it ab initio}, did not consider fine structure and
other coupling effects.

\begin{figure}[htb!]
 \centering
\includegraphics[width=8.3cm]{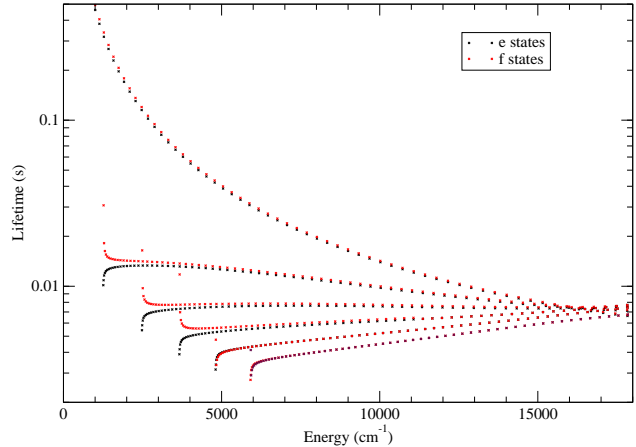}
\vspace{0.5cm}
 \caption{Lifetimes of the vibration-rotation states of $^{40}$CaH.
From the top
the ``curves'' are for vibrational state $v=0$, $v=1$ up to $v=5$.
Each vibrational curves is actually a doublet with the generally
longer-lived component representing the $f$-states in red and
the shorter-lived $e$-states being given in black.}
  \label{f:CaH}
\end{figure}

Figure~\ref{f:CaH} summarises our results. The figure shows a number
of features broadly in line with those seen for SiO above: for example
lifetimes for the $v=0$ levels again showing a $J^{-3}$ behaviour. For
higher $J$ values there appears to be little dependence on spin component.
However this is not true at low $J$ where it is found that the lifetimes
for the higher, $f$, component increase while those for the lower $e$ component
decrease significantly.

\subsection{AlO}

Aluminium monoxide is chosen not only because it is an important
astronomical species having been observed in sunspots
\cite{13SrViSh.AlO} and a variety of other stars
\cite{13KaScMe.AlO,04BaAsLa.AlO,12BaVaMa.AlO} but also for terrestrial
applications were emissions arise from rocket exhausts
\cite{96KnPiMu.AlO} and plasmas \cite{14SuPa.AlO}. Furthermore there
are direct experimental measurements of radiative lifetimes for
vibrational levels in the B~$^2\Sigma^+$ state
\cite{72JoCaBr.AlO,75DaCrZa.AlO}.

The ExoMol model for AlO considered the three lowest electronic states, all
of which are spin doublets: X~$^2\Sigma^+$, A~$^2\Pi$ and  B~$^2\Sigma^+$.
In practice the X and A states are close together and cross
at energies under consideration leading to significant interaction between
the states which is explicitly allowed for in the model used \cite{jt589}.
Lifetimes for AlO were computed using the linelist of \citeasnoun{jt598}
which considered $(v\leq 66,J\leq 300.5)$, $(v\leq 63,J\leq 300.5)$
and $(v\leq 40,J\leq 232.5)$ for the X, A and B states respectively.
This linelist is designed to be valid for temperatures up to 8000 K and considered
states as high as 40~000 cm$^{-1}$. However above 30~000 cm$^{-1}$ the line list
does not include all transitions meaning that for states in this region
the computed some lifetimes are artificially long; thus only states below 30~000 cm$^{-1}$ are considered.

\begin{figure}[htb!]
 \centering
\includegraphics[width=8.3cm]{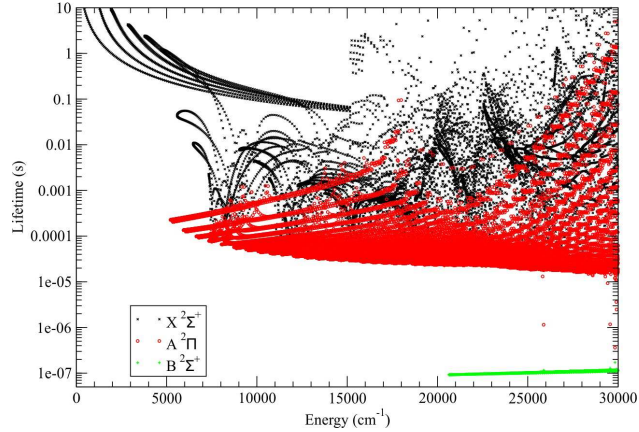}
\vspace{0.5cm}
 \caption{Overview of lifetimes for states of
 $^{27}$Al$^{16}$O.}
  \label{f:AlO}
\end{figure}

Figure~\ref{f:AlO} gives an overview of lifetimes of levels
belonging to the lowest three electronic states of $^{27}$Al$^{16}$O.
Broadly, the curves in the top left are associated with levels
belonging to the  X~$^2\Sigma^+$ ground state, the large clump
in the middle belong to the low-lying A~$^2\Pi$ state and the
short lived levels at higher energy belong to the B~$^2\Sigma^+$ state.

Superficially the structure of the  X~$^2\Sigma^+$  state lifetimes
are similar to those discussed above for SiO. However they differ in two
key aspects, firstly the states are much longer lived and secondly the shortest
lived curve is for levels with $v=0$ and lifetimes actually increase with
vibrational excitation. These features are due to the very flat dipole
moment curve of AlO which leads to unusually weak vibrational transitions
\cite{82LeLixx.AlO,jt598}.

\begin{figure}[htb!]
 \centering
\includegraphics[width=8.3cm]{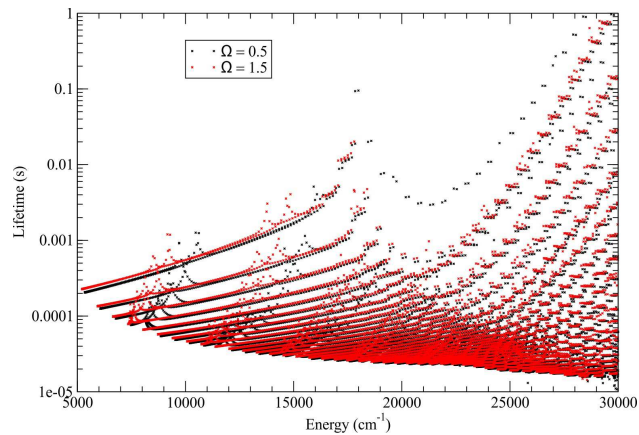}
\vspace{1cm}
 \caption{Lifetimes of the vibration-rotation states of
  A~$^2\Pi$ of AlO. }
  \label{f:AlOA}
\end{figure}

The lifetimes of levels associated with the A~$^2\Pi$ state vary by
many orders of magnitude. This is partly due to interaction with much
longer-lived levels from the X~$^2\Sigma^+$ state via spin-orbit
coupling; the mixing of the wavefunctions between the two states leads
to many structures in the lifetimes plot.  Furthermore the A~$^2\Pi$
state itself is split into $\Omega = \frac{1}{2}$ and $\Omega =
\frac{3}{2}$ states by spin-orbit coupling. As shown in
Fig.~\ref{f:AlOA}, the lifetimes of levels associated with these
curves are rather similar except when the levels interact with levels
associated with the X~$^2\Sigma^+$ state.

\begin{figure}[htb!]
 \centering
\includegraphics[width=8.3cm]{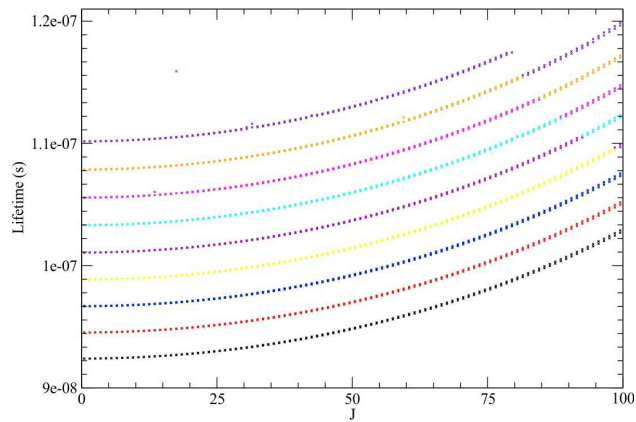}
\vspace{1cm}
 \caption{Lifetimes of the vibration-rotation states of
  B~$^2\Sigma^+$ of AlO. From bottom to top
the curves are for vibrational state $v=0$, $v=1$ up to $v=9$. The small
splittings visible at higher $J$s is due to spin-rotation effects.}
  \label{f:AlOB}
\end{figure}

 There available measurements of lifetimes for the $v=0$, 1 and 2
 levels of the B~$^2\Sigma^+$ state: \citeasnoun{72JoCaBr.AlO} obtained
 $128\pm6$, $125\pm5$ and $130\pm7$ ns respectively for these levels
 while \citeasnoun{75DaCrZa.AlO} obtained $100\pm7$, $102\pm7$ and
 $102\pm4$ ns respectively. These measurements were both rotationally
 unresolved for unspecified rotational distributions.
 Figure~\ref{f:AlOB} shows our rotationally resolved results for the
 lowest 10 vibrational states of the B~$^2\Sigma^+$ state. It is clear
 that our calculations reproduce the trend of slowly increasing lifetime
 with vibrational state. The calculations also suggest that the
 lifetimes increase with rotational excitation making our results
 agree with those of \citeasnoun{75DaCrZa.AlO} for $J$ in the region
 of 85 and those of \citeasnoun{72JoCaBr.AlO} for somewhat higher
 values of $J$.

\subsection{ScH}

\begin{figure}[htb!]
 \centering
\includegraphics[width=8.3cm]{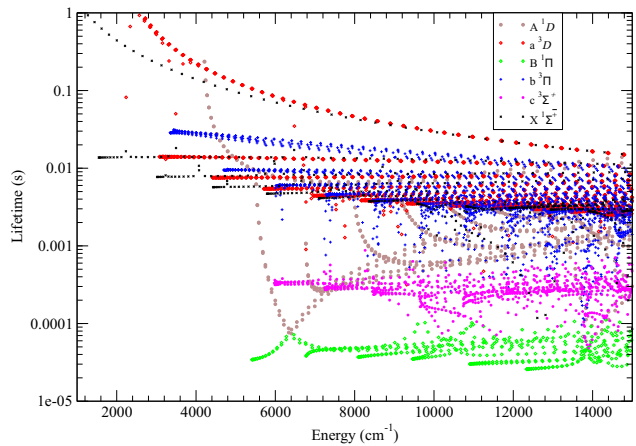}
\vspace{1cm}
 \caption{Lifetimes for states of
 $^{45}$ScH.}
  \label{f:ScH}
\end{figure}

Figure~\ref{f:ScH} shows scandium monohydride lifetimes calculated
using the linelist of \citeasnoun{jt599} which considered
the lowest six levels of ScH and $J\leq 59$. In energy order these states are the
 X~$^1\Sigma^+$ ground electronic state, a~$^3\Delta$, b~$^3\Pi$,
A~$^1\Delta$,  B~$^1\Pi$,
and  c~$^3\Sigma^+$. The minima of these states span a range of
little more than 6000 cm$^{-1}$. The many overlaps and couplings
between these states causes a complicated set of patterns in the lifetime
plot.

In contrast to the systems considered above, the model for ScH contains
electronic states with different spin symmetries. In particular, while
the ground state is a singlet, the first excited state is a triplet meaning
that there is no dipole connecting the two states. The a~$^3\Delta$
state is therefore metastable and its levels have long lifetimes.

Unlike the other systems considered here, the line list computed
by \citeasnoun{jt599} for ScH is completely {\it ab initio} since
there was insufficient available spectroscopic data to refine the
{\it ab initio} model. In addition, experience has shown that
present state-of-the-art
{\it ab initio} calculations give a much poorer starting point in
transition metal containing molecules. This means that our results
for ScH must be regarded as highly uncertain. Some experimentally
measured lifetimes for this system would be very helpful as input to
an improved model.

\subsection{Water}

In contrast to molecules like ScH and AlO, the intensity of key
astrophysical transitions for a molecule like water are both well
studied and well-known \cite{jt378,jt557}. However, as we show below,
the lifetimes of the various levels of water do not follow the simple
patterns of, say, SiO discussed above. The lifetimes are important for
models of the many water masers lines found both in ground
\cite{91NeMexx.maser} and excited \cite{12HiKiHo.maser} vibrational
states as they can affect the pumping routes and level populations
that lead to inversions. Indeed the water maser is a textbook example
\cite{12Gray.maser}. Similar considerations apply to fluorescence from
comets \cite{jt330,jt349}.

Here we use the BT2 line list of \citeasnoun{jt378} whose construction actually pre-dated
the ExoMol project. BT2 considered all states of water up to 30~000 cm$^{-1}$ with $J \leq 50$
and contains over 500 million transitions.

\begin{figure}[htb!]
 \centering
\includegraphics[width=8.3cm]{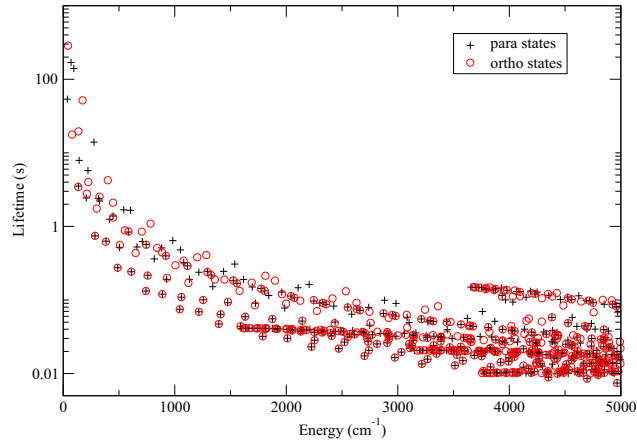}
\vspace{1cm}
 \caption{Lifetimes of the vibration-rotation states of H$_2$$^{16}$O.}
  \label{f:water}
\end{figure}

Figure~\ref{f:water} plots the lifetimes for levels lying below 5000
cm$^{-1}$; ortho and para levels are denoted separately since these
species effectively behave as separate molecules. While the general
structure of lifetimes is similar to that discussed for SiO, it is
clear that the rotational substructure introduces additional
complications. Indeed lifetimes for levels with the same total
rotational quantum, $J$, within the vibrational ground state can
differ by up to an order of magnitude.  While the shorter-lived
levels, which include the so-called water backbone \cite{12Gray.maser}
important for maser models, show some systematic structure in their
lifetimes and similar values for ortho and para species, the variation
between the lifetimes of the longer-lived levels is significant.  It
is the details of this structure which allow masing to occur: for
example when the lifetime of the higher level is significantly longer
relative to the lower, the upper level can overpopulate relative to
the lower, critical for masing to occur.

\section{Methane}

Methane is an important astronomical molecule and easily detected in
hot environments \cite{jt572}, but it does not possess a permanent dipole
moment hindering its detection in interstellar environments
\cite{91LaCaEv.CH4}. However, methane does possess a weak rotational
spectrum \cite{75CoHoxx.CH4,85OlAnBa.CH4,87HiLoCh.CH4,10BoPiRo.CH4} which is
important for studies of methane in giant planets and Titan
\cite{07WiOrOz.CH4}.  Lifetimes were computed using the 10to10 line
list of \citeasnoun{jt564} which contains almost 10 billion
vibration-rotation transitions with $J \leq 39$.

The high symmetry of CH$_4$ brings some unusual features to its
spectroscopic behaviour.  Nuclear spin statistics divides the levels
into three distinct groups conventionally known as ortho, meta and
para. One might expect that there would thus be 3 levels which do not
undergo dipole-allowed radiative decay corresponding to the lowest
level of each group. In fact, there are seven such levels. Labelling
them as $(J,n,$symmetry), where $n$ is simply a counting number,
they are the lowest ortho state (0,1,A$_1$)
at 0 cm$^{-1}$, the lowest meta level (1,1,F$_1$) at 10.48 cm$^{-1}$,
and the lowest para level (2,1,E) at 31.44 cm$^{-1}$, which are the
lowest three levels in the molecule. In addition the three ortho
states (3,1,A$_2$) at 62.88 cm$^{-1}$, (6,1,A$_2$) at 219.92 cm$^{-1}$
and (9,1,A$_1$) at 470.84 cm$^{-1}$ and the para state (4,1,E) at
104.78 cm$^{-1}$ do not have any dipole allowed transitions linking
them to lower levels. These states will probably decay slowly by other
means, such as via electric quadrupole transitions.

\begin{figure}[htb!]
 \centering
\includegraphics[width=8.3cm]{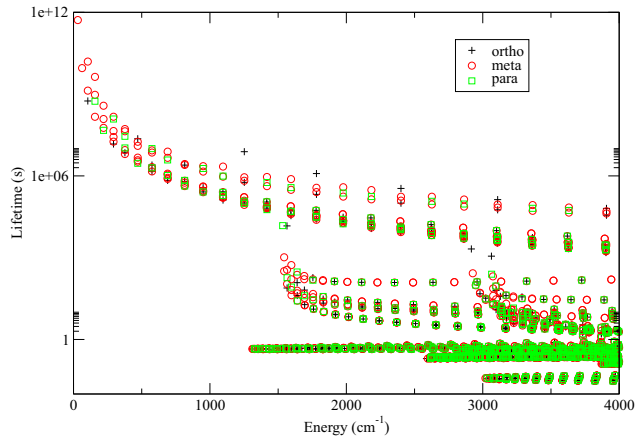}
\vspace{1cm}
 \caption{Lifetimes of the vibration-rotation states of $^{12}$CH$_4$.}
  \label{f:methane}
\end{figure}

Figure~\ref{f:methane} gives lifetimes of vibration-rotation states of
methane lying below 4000 cm$^{-1}$. The very weak nature of methane's
rotational spectrum means that states with no vibrational excitation
live for a long time and the figure actually spans 14 orders of
magnitude. In fact there are three very long-lived states which are
omitted from the figure. These are (3,1,F$_2$) at 62.88 cm$^{-1}$,
(6,1,A$_1$) at 219.95 cm$^{-1}$ and (12,1,A$_2$) at 817.10 cm$^{-1}$.
None of these states can undergo a dipole-allowed P ($\Delta J =-1$)
transition and instead each decay via a single, very long-wavelength Q
($\Delta J =0$) transition. Our calculated lifetimes are $1 \times
10^{23}$, $4 \times 10^{17}$, and $6 \times 10^{13}$ s, respectively.
These are huge: in particular the lifetime of the (3,1,F$_2$) is
predicted to be a thousand times the age of the Universe! Our precise
numbers are rather uncertain because the lifetime depends on the cube
of the wavelength which, for close-by energy levels, is calculated
with a significant uncertainty. Furthermore, given the long lifetimes,
it is likely that other non-dipole or two photon decays may become
dominant. However, the prediction that these three states are very
long-lived in collision-free environments is likely to be correct.
The very faint, pure rotational spectrum of methane 
has been studied by \citeasnoun{10BoPiRo.CH4}, who were able to determine
values for the very small dipoles linking the rotational states. It
is the small values of these transition dipoles combined with the small gap between
the states involved that leads to the exceptionally long radiative lifetimes for
certain states.

\section{Cooling functions}

\begin{figure}[htb!]
 \centering
\includegraphics[width=8.3cm]{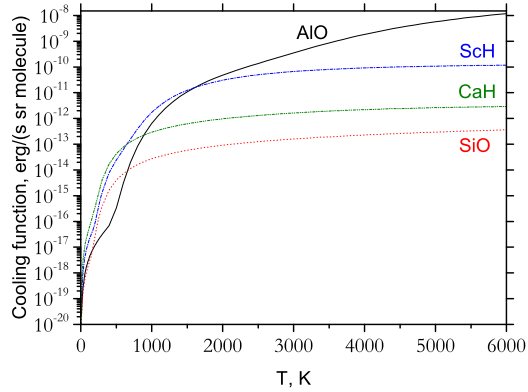}
 \caption{Temperature-dependent cooling functions for $^{40}$CaH, $^{28}$Si$^{16}$O, 
$^{27}$Al$^{16}$O and $^{45}$ScH.}
  \label{f:cool}
\end{figure}

\begin{figure}[htb!]
 \centering
\includegraphics[width=8.3cm]{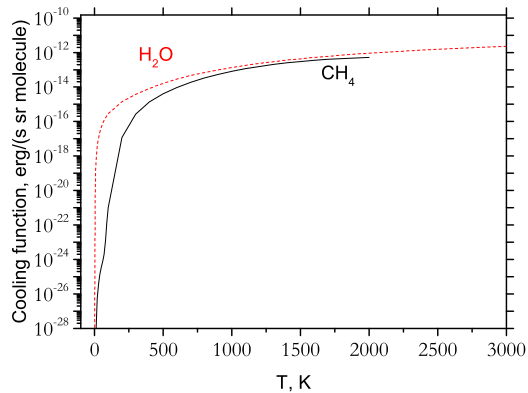}
 \caption{Temperature-dependent cooling functions for methane and water.}
  \label{f:cool2}
\end{figure}

Cooling functions were computed for the molecules discussed above
using the same line lists.  Results are given here graphically, see
Fig.~\ref{f:cool} and \ref{f:cool2}, and are given numerically in 1 K
intervals on the ExoMol website. The cooling functions of AlO and ScH
show structure at lower temperatures due to the presence of
transitions from exited electronic states. The strong electronic
transitions also lead to the cooling functions for these two molecules 
being significantly stronger, especially of AlO at $Т>2000$~К.  The
cooling function for water was computed up to $T$=3000~K, which is the
temperature limit of the BT2 linelist used \cite{jt378} and that for
methane up to $T$=1500~K, the limit for the 10to10 list. Methane is
actually a rather inefficient cooler below about 500~K which can be
understood in terms of the long lifetimes of the rotationally excited
states discussed above.

It should be noted that the cooling function of CaH is expected to underestimate at  $T>2000$~K) 
due to the absence of the  excited electronic state in the ExoMol line list \cite{jt529}.
 In general at high temperatures the cooling functions will be too low, due to the incompleteness 
of the underlying line lists. 
However it is possible to give an improved estimate, $W_{\rm corr}(T)$,
using the partition function, if available:
\begin{equation}
 W_{\rm corr}(T) = W(T) \frac{Q(T)}{Q_{\rm LL}(T)}
\end{equation}
where $Q_{\rm LL}(T)$ is the partition function obtained by summing over the energy
levels used to construct the cooling function, and $Q(T)$ is converged high-temperature
partition function. Good high-temperature partition functions  are
available for certain key molecules such as H$_3^+$ \cite{jt169}, water \cite{jt263},
ammonia and phosphine \cite{jt571}, and methane \cite{08WeChBo.CH4}; the compilations made for
HITRAN have also proved to be reliable up to 3000 K \cite{00GaKeHa.partfunc,11LaGaLa.partfunc}.
Thus, for example, Eq.~(4) should probably be used to scale the cooling function given
for methane in Fig.~\ref{f:cool2} since the 10to10 line list used to construct this function
is incomplete for $1500 < T < 2000$~K.

\section{Discussion and Conclusion}

We have shown that by using extensive spectroscopic line lists to
compute state-dependent lifetimes, interesting and underlying
behaviour of the radiative properties of molecules can be revealed.
The ExoMol project is systematically constructing line lists for a
variety of astrophysically important molecules. The ExoMol format has
recently been extended \cite{jtexo} so that the states file can be 
used to capture information about each state in the molecule; this 
includes its computed lifetime. Lifetime data for the molecules
considered here can be found on the ExoMol website (www.exomol.com)
and lifetime data for other molecules will be placed there as it
becomes available. In particular, the importance of lifetime cooling
effects for H$_3^+$ were discussed in the introduction.  However,
despite there being an extensive H$_3^+$ linelist being available
\cite{jt181}, we have chosen not to present data for this system. For
technical reasons the available linelist, despite its proven accuracy
\cite{jt512}, does not include the weak, ``forbidden'', far infrared
pure rotational transitions which govern both lifetime effects for
states in the vibrational ground state and the cooling function at low
temperatures. A new, more complete linelist for H$_3^+$ is currently
being computed as part of the ExoMol project \cite{jtH3+}. This new
linelist will be used to both calculate state-dependent 
lifetimes and a new cooling curve.

Lifetimes also provide an important link with laboratory experiments.
Many astronomically important species and/or transitions can only be
reproduced in the laboratory in non-equilibrium conditions.  This
makes it very difficult to measure transition intensities. The
alternative approach of measuring lifetimes has the advantage that it
does not depend on the level populations or thermaisation of the
sample being studied. More of such measurements would be very helpful
for comparison and validation of our calculations.


We have shown that the presence of complex electronic spectra is
responsible for a non-monotonic behaviour of $W(T)$ at low
temperatures and strong cooling at high temperatures.  We will
generate cooling functions routinely on the grid of 1~K and provide
them together with the partition functions as part of the ExoMol
database.

\section*{Acknowledgements}

This work was supported by the ERC under the Advanced Investigator Project
267219.

\section*{References}
 \bibliographystyle{jphysicsB}

\end{document}